\documentclass{article}
\usepackage{graphicx}
\usepackage[usenames,dvipsnames]{xcolor}
\usepackage{hyperref}
\hypersetup{
	pdfcreator={LaTeX with abnTeX2},
	pdfkeywords={abnt}{latex}{abntex}{USPSC}{trabalho acadêmico}, 
	colorlinks=true,       		
	linkcolor=blue,          	
	citecolor=blue,        		
	filecolor=magenta,      		
	urlcolor=blue,
	allbordercolors=black,
	bookmarksdepth=4
}
\usepackage[utf8]{inputenc}
\newcommand{\ttt}[1] {
	\texttt{<#1>}}
\newcommand{\tttt}[1]{\texttt{#1}}

\begin{document}
\title{An anthropological account of the Vim text editor:\\
features and tweaks after 10 years of usage}
\author{Renato Fabbri\\
\texttt{renato.fabbri@gmail.com}\\
University of São Paulo,\\
Institute of Mathematical and Computer Sciences\\
São Carlos, SP, Brazil
}
\maketitle
\begin{abstract}
The Vim text editor is very rich in capabilities
and thus complex.
This article is a description of Vim
and a set of considerations about its usage and design.
It results from more than ten years of experience
in using Vim for writing and editing various types of documents,
e.g. Python, C++, JavaScript, ChucK programs;
\LaTeX, Markdown, HTML, RDF, Make and other markup files;  
binary files.
It is commonplace, in the Vim users and developers communities,
to say that it takes about ten years to master (or start mastering)
this text editor, and I find that other experienced users
have a different view of Vim and that they use a different
set of features.
Therefore, this document exposes my understandings in order
to confront my usage with that of other Vim users.
Another goal is
to make available a reference document with which new users
can grasp a sound overview by reading it and the discussions that
it might generate.
Also, it should be useful for users of any degree of experience,
including me, as a compendium of commands, namespaces and tweaks.
Upon feedback, and maturing of my Vim usage,
this document might be enhanced and expanded.
\end{abstract}
{\bf keywords:} Vim, Text editor, Anthropological computer science,
HCI, Tutorial

\section{Introduction}\label{intro}
Vim is a very complex editor,
most often considered to be
matched only by Emacs.
They both are the standard advanced text editors
of the free software and open source communities
and have been developed for decades.
Vim is very useful because:
\begin{itemize}
  \item it is meant to be a plain text (e.g. ASCII, UTF-8) editor
  and does not (by standard) insert special characters (e.g. for
    formatting or with binary instructions).
  \item It has a powerful architecture and set of commands.
  \item It is highly configurable and most often the users
  hold a set of commands for standard settings and routines kept in
    the vimrc~\cite{vimrc} and other configuration files.
  \item It has been used for more than 25 years, is based on
    established technologies, and is an established text editor.
    Thus, it has a considerable and seasoned user
    base, high quality (both official and unofficial) documentations, 
    and countless publicly available scripts, most often in the form of
    plugins.
\end{itemize}

This document describes the Vim text editor
and proposes a set of sharpening of the user
experience through simple tweaks
and utilization strategies.
The contents herein presented is a
report on the overall understandings I
have of Vim after a bit more
than ten years using it,
resorting to the (very mature) official documentation
whenever possible.
The purposes of this document are:
\begin{itemize}
  \item to help new users grasp Vim essentials
  and convenient practices.
  \item To attain a sound overall description of the editor.
  \item To record the comprehension about the editor that
  a user (me) has after $10$ years of usage.
  \item To confront my usage with that of other experienced
  users. This is helpful for me, but also for the other users
  as they might benefit from this content and from discussions
  that might arise from it.
  \item To propose some enhancements to Vim through simple tweaks and potential plugins.
\end{itemize}

Advanced users might just skim through
Section~\ref{basics}, where standard capabilities of
the editor should become clear, and consider more carefully
Section~\ref{namespaces}.
The concluding remarks and proposed enhancements in Section~\ref{conc}
may also deserve some attention because
Vim is constantly evolving
and there are many possible enhancements
(often made available e.g. as plugins).

\subsection{Further remarks about this document}
This document is written is a DRY KISS
(Don't Repeat Yourself, Keep It Simple Stupid) style.
Complex is to master the use of Vim
and one finds sound
references in help files and a nice vimrc.
Therefore the following content is kept
as uncomplicated and original as possible.
Also, because of Vim's complexity and entailed
bond of this document to my usage,
there is an anthropological component
which is evident in the occasional use of the first person.
This can be understood as anthropological computer science~\cite{anPh,anPh2}
and considered to help
in the technological groundwork of the civil society.
Accordingly, here are some notes about my experience:
I've had experience with other editors, e.g. Kate, gedit, and Notepad2.
I used Vim for writing and editing computer code 
(Python, Javascript, C++, ChucK, bash, etc), markup languages  
(HTML, CSS, RDF, Markdown, \LaTeX, etc) and binary files.
Eventually, I edited database files and other types of files.
Within Vim, I mostly write (web and scientific) software,
scientific articles, music, poems and short stories.

\subsection{Historical note}
Vim was first released publicly in 1991.
It is a cross-platform GNU licensed free and open source
extended clone of Bill Joy's vi text editor.
Vi was written in 1976 as a hard link to ex: a shorthand
to start ex in visual mode, i.e. vi is ex.
Vim's development is coordinated (and performed) since the beginning
mainly by Bram Moolenaar.
Today, current bleeding-edge version is 8.0.1401.
I found no explicit stable, alpha or beta versions.
I found no scientific article on Vim 
(this might be the first one), although there are
books, software and third-party documentation on the
web.~\cite{wikiVim}

\section{Basics}\label{basics}
Vim's interface is text-based.
In the GUI (gVim),
there are convenient menus and toolbars
but all functionalities are still available though
the command line mode.
Vimscript is the internal language of Vim,
and is often used for scripting by users
although other languages might be used 
(e.g. Python, Perl, Lua, Racker, Ruby and Tcl). 
Each line of a Vimscript is a command on the
command-line mode.
This section may be considered a tutorial
that focuses on the namespaces, i.e. sets of tokens
that carry values or triggers procedures.
One should see Appendix~\ref{sec:not}
to understand the notation Vim uses.

\subsection{The bare minimum}\label{minimum}
You open a file at Vim startup by executing
the command: \texttt{vim <filename>}.
Inside Vim, you start in the normal
mode, and might want to move around using
\texttt{h-j-k-l} for left-down-up-right.
To insert characters, move your
cursor to the desired location an press i,
which puts Vim in the insert mode.
Go back to normal mode by pressing
\texttt{<ESC>}, \ttt{C-$[$} or \texttt{<C-C>}.
You save the file by typing \texttt{:w<CR>},
and exit Vim by typing \texttt{:q<CR>}.
You can save and quit with \tttt{:wq<CR>}
or \tttt{:x<CR>} or \tttt{ZZ}.

\subsection{Vim help}
Help on using Vim is found in various places.
The standard resource is the Vim help files.
They are accessed by typing \texttt{:h <anything><CR>}
in normal mode.
Examples of such \texttt{<anything>} are:
color, navigation, :vs, vimtutor.
Type 
\texttt{:h usr\_toc<CR>}
to access the official User Manual,
which is considerably lengthy and complex
and is usually not read by users before a few years
of experience.
In learning Vim, one
might want to run the \texttt{vimtutor} command
(outside Vim) to start the Vim Tutor.

There are good resources on the Web for learning
and tweaking Vim:
\begin{itemize}
  \item ``Vim Adventures'' is an online RPG game for practicing
  and memorizing Vim commands. This game is quite famous among Vim users.
  \item There are official and semi-official Vim sites e.g.:
  \url{www.vim.org}, \url{https://www.vi-improved.org} and
  \url{http://vim.wikia.com/}.
  \item Many hacks, understandings and general issues
  (e.g. how to make such a move) are asked and answered
  in online platforms (e.g. Quora, Stack Overflow, Stack Exchange, Reddit, Email list, IRC Channel).
  One often finds these links through a search engine.
  \item Many videos about Vim are publicly available.
  One traditional site is \url{http://vimcasts.org},
  but you might find them using a search engine
  (e.g. \url{http://derekwyatt.org/vim/tutorials/}) or in Youtube and Vimeo.
\end{itemize}

\subsection{Namespaces}\label{namespaces}
Vim is a text editor ouroboros~\cite{ouroWiki} because
text and writing alters text and writing.
What you end with is a collection of namespaces where tokens have scalar
or arbitrarily complex values.

\subsubsection{Commands and mappings}
There are commands, i.e. typing sequences which trigger automated actions,
for each mode:
\begin{itemize}
	\item in Normal mode all keys are mapped to commands.
		There is redundancy and additional commands
		using Ctrl and Shift keys.
		Some keys expect a second key,
    and have combinations not used (thus available for new mappings),
		specially the z and g.
    See Section~\ref{normal} and Appendix~\ref{notes}
    for more insights into the commands available in the
    normal mode.
	\item In the other modes, the sequences available for mappings are more obvious and abundant.
		One should look at \tttt{:h index} to know about all the standard mappings
		and use \tttt{:map} to list the user-defined mappings.
\end{itemize}

\ttt{C-\textbackslash} is often reserved for extensions,
which makes it a safe namespace to use (while there are no
such extensions).

A colon command can be written as a string
and executed by the \tttt{:execute} colon command.
E.g. \tttt{:execute 'vs afile.txt'}.
As there are colon commands that execute commands in other
modes, e.g. \tttt{:normal ?\textasciicircum def},
the \tttt{:execute} is a way to build commands in any mode,
e.g.
\\\tttt{:execute 'normal i' . string(atan(bufnr('\%')))}.

\subsubsection{Variables}
There are some types of variables in Vim:
\begin{itemize}
	\item Environment variables: names start with \$ and hold system
		variables, such as \tttt{\$PATH} and \tttt{\$PWD}.
	\item Option variables: names start with \& and are meant to control the behavior of the editor.
		One might change a value through set or let, e.g.
		\tttt{:set bg=light} or \tttt{:let \&bg=light}.
	\item Registers: start with @ and are meant for automation and transfer of texts (copy and paste).
    More on register in Section~\ref{state} and scattered in this
    document.
	\item Internal variables are created with let and preferably have a prefix:
	\tttt{b:}, \tttt{w:}, \tttt{t:}, \tttt{l:}, \tttt{s:},
		are local to the buffer, window, tab page, function, and
		sourced Vim file, respectively.
 \tttt{v:}, \tttt{g:} are global, the first are predefined by Vim.
		\tttt{a:} is for function arguments.
		If there is no prefix, the variable is global or internal to a function if occurring inside a function.
		More about internal variables in \tttt{:h internal-variables}.
	\item The value of a variable can be a scalar, string, list, dictionary, function reference, etc (see \tttt{:h eval}).
\end{itemize}
You can echo any of such variables or use them in expressions.
Notice that you will only be able to echo a \tttt{b:} variable inside
the buffer where it is defined.
For all the Vimscript capabilities, including loops, conditionals,
and builtin functions, refer to \tttt{:h vim-script} and Section~\ref{script}.
Classes are possible only in rudimentary forms, e.g. through dictionaries,
but the language is otherwise overall quite powerful,
specially in dealing with text and editor behavior, as expected.

\subsubsection{State lists}\label{state}
Vim keeps a number of useful lists which expresses the state of the editor:
\begin{itemize}
	\item All the entered commands are accessed through \tttt{:hist a}. 
    The tokens ``\tttt{a / e : i d}'' may be used for specific types of commands, such
		as search and colon commands.
	\item File buffers are kept with numeric ids. See buffers with \tttt{:ls} and load a buffer to the window with \tttt{:b <num or token in file name>}.
  \item The windows open are listed in \tttt{:ls} with a character \tttt{a} in the second column, and are listed with \tttt{:tabs}.
	\item Tabs list can be reached through \tttt{:tabs}.
		It is usual both to show and hide the tabs bar (mapping
    in~\cite{vimrc}, discussion in Section~\ref{visual1}).
	\item Jumps are available through \tttt{:jumps}.
		One positions the cursor at each jump through \ttt{C-O} and \ttt{C-I}.
  \item Registers are available through the \tttt{:reg} command, as variables and through shortcuts in different modes.
    They also keep track or your copy, edition and deletion and are promptly defined
    by recording a typing sequence with the \tttt{q} normal command.
    Vim keeps only the last edition, in register \tttt{".}.
    An autocommand to keep the four latest inserts is in~\cite{vimrc}.
    A hack to keep the latest deletions and copies in the standard register
    might follow the same pattern, but use another \tttt{:h event} and
    monitor register \tttt{""} (maybe also monitor \tttt{"0}).
	\item An undo list can be accessed through \tttt{:changes}.
	\item A list with all the sourced scripts in a Vim instance is displayed through \tttt{:scriptnames}.
	\item The markers defined are listed with the \tttt{:marks} command.
		These are set by \tttt{mX} in normal mode, where X is the marker identifier.
		Uppercase letters are cross buffers.
	\item Quickfix and Location lists which are populated through \tttt{:vim} and \tttt{:make}
		and variations, such as \tttt{:grep}.
		One might run \tttt{:vim /section/ \%} and then \tttt{:copen}
		to open the Quickfix window, where the lines of occurrence are in sequence
		and one can \ttt{CR} one of them to have the cursor in the main window active at
		the first character of the match.
		One might run \tttt{:lvim /section/ \%} and then \tttt{:lopen}
		to use the location-list window instead of the Quickfix, which
		is very similar, but one per window instead of one per buffer.
		More information in \tttt{:h quickfix}.
	\item A tags list have to be made so one can use tags.
		Most often one will generate the tags list using the exuberant ctags,
		which supports dozens of languages.
		E.g. \tttt{:!ctags-exuberant functions.py} or \tttt{!ctags-exuberant -R ./},
		and then using \ttt{C-]} to
		go to the position of tag under cursor, and \tttt{:vs tags} to open
		the tags file.
	\item The argument list holds a list of files to be edited or browsed. 
		The list can be input at Vim startup (e.g. \tttt{\$ vim file1.txt file2.py})
	or using commands (e.g. \tttt{:ar ./*}).
		The file being edited is changed by \tttt{:n} and \tttt{:p} commands,
		one might perform actions on each file in argument list using \tttt{:argdo}.
		All files in argument list are also in the buffers list.
    For further information, see \tttt{:h arglist} and Section~\ref{netrw}.
  \item All the autocommands might be accessed through \tttt{:au}.
    They are event-triggered actions Vim performs.
\end{itemize}

A file with information about the state of the editor
can be achieved through: \tttt{:source \$VIMRUNTIME/bugreport.vim}.
Also, this script might be examined because it has
a collection of commands to access various settings of Vim.
Another good list of commands to know about Vim's state is kept on
\url{http://vim.wikia.com/wiki/Displaying_the_current_Vim_environment}.
I would specially highlight the \tttt{:syntax} command because
it displays token groups and their meanings when run
inside e.g. a \tttt{.py}, \tttt{.vim} or help file.

\subsubsection{On the persistence of visual cues about the editor state}\label{visual1}
You can keep track of the editor state though commands, as stated
above.
Also, one might rely on persistent visual
cues, specially the tabs bar, the status line, and the line
reserved for the command-line.
A good strategy I find is to have selective visual cues of the state to make persistent or hide
and a mapping to toggle each of them.
Currently, I toggle byobu/screen/tmux bar with \ttt{F5},
status line and tabs bar with \ttt{localleader-T or B} according to script~\cite{vimrc}.
I am mostly using the cleanest setting, toggling on the tabs bar and status line sometimes.
Numbering is always there, I rarely turn them off but keep the mappings \ttt{leader-n or N} to toggle just in case.
Instead of keeping the status bar, I use \ttt{C-G} to know about the
file and \ttt{gC-G} to know more and rarely.
It seems not possible to remove the statusbar between horizontal splits.
After asking in online forums and experimenting, I realized that it seems
reasonable to keep at least one line dividing the windows, so
if it comes to it, I just \tttt{set statusline=-}.
Unfortunately, as far as I could dig, one will need to alter
the Vim's source code to enable a horizontal split without losing a line.
For me, it would be ideal for this feature to have a visual cue of the
first and/or last line of the windows in the lines-number
column, or complete the spaces and empty chars with \$\$\$\$ or so.

\subsubsection{On the persistence of the editor state}
For state persistence, one might keep an undo file for each file as
in~\cite{vimrc}.
Sessions are easy to manage, enabling one to save and load the
editor's state, with the opened windows, tabs, buffers, etc.
The mappings in~\cite{vimrc} keep the sessions
in a reasonable directory and makes it easier to remember and tweak
the standard commands to deal with sessions.
More information in \tttt{:h sessions}.
One might use \tttt{:h views} to keep the state of one window,
but sessions keep all the states from all windows.
This entails a strategy to deal with Vim that is similar to
the use of
Byobu/Tmux/Screen\footnote{\url{https://en.wikipedia.org/wiki/Byobu_(software)},
\url{https://github.com/tmux/tmux/wiki},
\url{https://www.gnu.org/software/screen/}},
because one can rely on restoring the state of the windows.
The main limitation I found to this approach is that
Vim is not keeping track of the terminals opened.
If you open a terminal inside Vim with the \tttt{:term}
command, you will save the session as usual, but when loading
you get dummy empty windows for them and an error message.
More about Terminal-Job mode in Section~\ref{terminal}.

Autocommands are the standard way to define event-triggered routines in Vim.
These are often related to particular file types, but are also often
in defining the automated behavior of Vim.
If the autocommands are placed inside a configuration file (e.g. the
vimrc), the automated behavior is persistent across Vim instances.
E.g. in~\cite{vimrc} is found an autocommand for keeping track of the last inserted texts in the \tttt{".lkjh} registers (\tttt{@.lkjh} variables).

\subsection{Using Vim's modes}\label{modes}
These are the  basic and fully implemented modes of usage in Vim:
\begin{itemize}
  \item Normal mode: used for changing
  the position of the cursor or the text displayed
  at the window.
  A core goal of the normal mode it to support fast
  navigation of the document while allowing
  the typist to maintain the fingers on the home row
  (i.e. on the center of the keyboard).
  The mode is also used for manipulating text
  (e.g. copy, paste, delete, change case) and
    changing to other modes.
    More in Section~\ref{normal}.
  \item Insert mode: for inserting text. More in
    Section~\ref{sec:ins}.
  \item Command-line mode: for entering Ex commands. More in
    Section~\ref{sec:com}.
  \item Ex mode: similar to command-line mode,
  but more specialized for running various Ex commands.
    More in Section~\ref{sec:ex}.
  \item Visual mode: for making, manipulating and navigating
  selections of texts.
  \item Select mode: similar to visual mode but
  favors CUA\footnote{IBM Common User Access:
    \url{https://en.wikipedia.org/wiki/IBM_Common_User_Access}.}.
\end{itemize}

There is another basic mode, but it is not fully implemented:
the Terminal-Job mode (more in Section~\ref{terminal}.
There are seven additional modes which are mostly subordinate 
to the basic modes and that will be described when convenient.
The manual page for Vim modes can be accessed by typing
\texttt{:h vim-mode}.
Some of the modes are now further considered for
the achievement of
an overview of the Vim usage possibilities.

\subsubsection{Normal mode}\label{normal}
Sometimes also called navigation or command mode,
the normal mode is most powerful for
navigating, manipulating texts and changing to other modes.
The simplest of these three is changing to other modes:
type any of these letters to change to insert mode:
\texttt{iIaAoOsScC}. More on the transition between
normal and insert mode on Section~\ref{navIn}.
Type any of these characters to change to command-line mode:
\texttt{:/?}.
Type \texttt{Q} to enter Ex mode.
Type \texttt{v}, \texttt{V}, or \texttt{CTRL+V} to enter visual mode.

For very basic and naive navigation, one should check Section~\ref{minimum}.
There are many facilities to navigate Vim
as explained in \texttt{:h navigation}.
Most often, one uses:
\begin{itemize}
  \item \texttt{Ctrl+(d,u,f,b)} for half-page down and up
and whole page down and up, respectively, although these commands might
be set to scroll a different number of lines.
  \item \texttt{Ctrl+(e,y)} to move the window one line down or up.
  \item \texttt{(w,b,e)} to move to the next and previous word,
    and next end-of-word.
    There are motions to iterate over sequences of characters separated
    by special characters (e.g. punctuation and parenthesis) as
    specified by the output of \texttt{:se iskeyword}.
    To iterate over space-separated tokens, use \texttt{W,B,E}).
    To move to the end of last word, one might use \texttt{be}
    or \texttt{ge}.
The \texttt{),(} commands iterate through sentences,
		\tttt{\},\{} through text blocks separated by empty lines.
  \item \texttt{(fX,tX,FX, TX)} to move to or just before any X character,
	  \tttt{;} and \tttt{,} for next and previous found character.
  \item Search with \tttt{/} or \tttt{?}, although these are in truth command-line commands.
  \item \texttt{CTRL+(o,i)} to move to an older or newer position in
    jump list.
  \item \texttt{'X,`X} to move the cursor to a mark bind to the alphanumeric character X:
    \texttt{`X} moves to the exact position while \texttt{'X} moves to
    the first non-blank character of the line.
    A mark is registered by the user in any cursor position
    by typing \texttt{mX}, where X is any letter.
    If X is lowercase, the mark is local to the buffer (the file),
    if it is uppercase or numeric, it is global to the Vim session
    (cross buffers).
\end{itemize}

For changing the text, usual commands include:
\begin{itemize}
  \item \texttt{d\{motion\}} to delete the characters
    involved in the motion command.
  \item \texttt{dd,D} to delete a line or from the cursor to the end of the line.
  \item \texttt{x,X} to delete the character under or before the cursor.
  \item \texttt{$\sim$} to swap the case of a character or selection.
  \item \texttt{gu\{motion\},gU\{motion\},g$\sim$\{motion\}} to make lowercase, uppercase or switch the case of the characters involved in the motion.
\end{itemize}

There are way more commands to change the texts.
Some of them are discussed in Section~\ref{navIn}
because they involve a transition to the insert mode.
A thorough consideration of the commands in the normal mode
is found by executing \texttt{:h navigation}, 
\texttt{:h change.txt}, \tttt{:h index}.

\subsubsection{Insert mode}\label{sec:ins}
Once in the insert mode, the character keys
input the characters at the cursor position at the current buffer.
One can exit insert mode by pressing \texttt{<Esc>} (or \ttt{C-$[$} or
\ttt{C-C}),
and Vim will be put in the normal mode.
Most useful commands in insert mode include:
\begin{itemize}
  \item \ttt{C-O} to execute one and only command in normal mode.
	  This enters a secondary mode (see Section~\ref{modes})
  \item \ttt{C-R} to paste a register (a variable starting with
    '@', defined, copied or recorded through a \tttt{q} command in
    normal mode and as covered in Section~\ref{state}).
  \item \ttt{C-T} to indent current line.
  \item \ttt{C-U,W} to delete all chars from cursor to the beginning of
    the line or to the previous word.
  \item \ttt{C-N,P} to find next and previous keywords that match the prefix at hand.
  \item \ttt{C-X} commands for scrolling the window with multiple
    \ttt{C-E} and \ttt{C-Y} strokes and for some completion facilities.
\end{itemize}

\subsubsection{Normal $\rightarrow$ insert modes}\label{navIn}
Many commands bridge from Normal to Insert modes,
e.g. \tttt{iws} or any of these letters: \tttt{csrCSR}.
These make convenient the replacement of text and populates registers.
The absence of a short command to insert one char is
a known issue in Vim.
Reasonable mappings to insert or append a char to and around
another char are in~\cite{vimrc}.
Vim couples operator and motion commands by design.
There are many operator commands that take the editor from
the normal mode to the insert mode, most of them
favoring deletion or change, as detailed in \tttt{:h operator}.
Motion commands are described in \tttt{:h motion}.

\subsubsection{Command-line mode}\label{sec:com}
This mode is dedicated to writing colon, search and filter commands,
entered through typing \tttt{:}, \tttt{?}, \tttt{/} and \tttt{!} in normal or visual modes.
Most useful commands in this mode include:
\begin{itemize}
  \item \ttt{C-B} and \ttt{C-E} to move cursor to the beginning and
    end of the line.
  \item \ttt{C-W} and \ttt{C-U} to delete last word or everything
    until the cursor.
  \item \ttt{C-R} to paste a register (as in insert mode).
\end{itemize}

\subsubsection{Ex mode}\label{sec:ex}
One might use the normal commands \tttt{q(:,/,?)} to have a window
with the colon or search command history to be edited normally,
and the chosen command can be run with \ttt{CR}.
Vimscript was largely based on the ex editor~\cite{ex},
and a more advanced user might use it for prototyping
by defining mappings and settings and managing scripts
probably in a \tttt{plugin/} folder of a directory in \tttt{:echo \&runtimepath}.
In the default interface started with the \tttt{Q} command in normal mode,
each command is input without entering : again. Use \tttt{:vi} to exit Ex mode,
follow documentation from \tttt{:h Ex-mode} for further information.
I might be loosing something, but for
tweaking I use the command-line window accessed through \tttt{q:}.
I find it comfortable to browse the history in a normal-like mode,
edit them also using the insert mode as usual,
and having auto-completion when pressing \ttt{TAB}.
None of such features are available in Ex mode by default
and
customization by mappings have to be performed through autocommands as in~\cite{vimrc}.

\subsubsection{Terminal-Job mode}\label{terminal}
This mode is reported as not having reached a stable usage design
(see \tttt{:h terminal}).
I find that it works exceptionally well and have used it to run
scripts in an IPython shell, compile latex files, and even open PDFs and images.
Vim browsing of windows and text manipulation is well developed,
so the Terminal-Job mode enables a very convenient integration
of files being edited and bash terminals,
more traditionally achieved through Byobu/Tmux/Screen terminal
multiplexers.
Most useful commands in the terminal-Job mode include:
\begin{itemize}
  \item \tttt{<C-W>\_N} and \tttt{<C-W>\_:} for entering the Terminal-Normal
    and command-line modes. Terminal-Normal mode is very similar to
    Normal mode, but one cannot change the text, cannot enter insert
    mode, and the status line reports if the job is finished or not.
  \item \tttt{<C-W>\_"} to paste a register.
  \item \tttt{i} for entering the Terminal-Job mode from the normal mode.
\end{itemize}
\noindent It is useful to define the same mappings for navigating splits and tabs
for both Normal and Terminal-Job modes, as in~\cite{vimrc}.

Because browsing the interface in Vim is fast,
and it is very comfortable to copy the terminal lines
in the Terminal-Normal mode,
it is often handy to keep (e.g. a tab with) some terminals:
e.g. one with an IPython shell, another two for compiling \LaTeX\
and opening PDF files (e.g. with \tttt{\$ evince <filename.pdf>}).
There are even more convenient ways to use the terminal inside Vim.
For example, 
one might use \tttt{:term} with the \tttt{++hidden} and \tttt{++close}
options to compile a \LaTeX\ file in the background without needing to further manage the
terminal and in a non-blocking manner: \tttt{:term ++hidden pdflatex
\% \#}.
I found the mappings on~\cite{vimrc} very helpful for directing
editor focus to splits (\ttt{C-hjkl}) and tabs (\tttt{gr}, \tttt{gt}),
which I make available across Terminal-Job and Normal modes.
But I've been thinking on using \ttt{C-} commands also
for tabs (not only for splits), both to avoid mixing typing sequences
with commands, and to allow the use of the commands in all Normal,
Insert and Terminal modes.

\subsection{Netrw}\label{netrw}
The standard interface of Vim for browsing file trees is Netrw.
It starts when you open a directory, such as with \tttt{:e .<CR>}.
It has solid support for browsing remote file trees (such as over ssh or
ftp) and handy e.g. mappings to open the files as splits and tabs
(specially \tttt{pot}).
Most useful commands in netrw include:
\begin{itemize}
  \item \tttt{d} and \tttt{\%} for creating directories and files.
    \ttt{Delete} removes both.
  \item \tttt{mf} and \tttt{mb} for marking files and bookmarking directories.
  \item \tttt{gb} and \tttt{uU} are used to load directories
    while marked files might be copied, moved, edited, grep-ed, tagged and migrated
    to and from the argslist as in \tttt{:h netrw-mf}.
\end{itemize}

There is no insert mode in netrw interface;
the commands in \tttt{:h netrw-explore} are
convenient for opening the directory of the
file being edited;
further information is in \tttt{:h netrw}.

\subsection{Standard configuration files and directories and my .vim/vimrc}
You can check the scripts Vim loads by using the debug script mentioned
in Section~\ref{state}.
By default, \tttt{$\sim$/.vimrc} and \tttt{$\sim$/.vim/vimrc} files are run by Vim at the beginning of the startup.
One might edit the vimrc file with \tttt{:e \$MYVIMRC}
and reload it with \tttt{:source \$MYVIMRC}.
Mappings in~\cite{vimrc} include such commands
to encourage a continuous enhancement of Vim settings
(they have helped me to improve my settings without unnecessary 
hassle and fast).
Any other file might be included to run at startup by
adding a line \tttt{:source <afile.vim>} in vimrc.
In fact, it is on vimrc that one usually specifies the plugins
and plugin managers they use.
Use an \tttt{after/} folder of a directory in \tttt{:se runtimepath}
or follow some patterns described on the next section to change the
scripts and sequence of them to be loaded.
The vimrc from other users are most useful for one to comprehend
and pick convenient practices and settings.
In fact, a vimrc file is most often a collage of excerpts of vimrc
files from other users.

\subsection{Plugins and packages}
One can see the list of the standard plugins
with \tttt{:h standard-plugin-list} command.
Any \tttt{plugin/**/*.vim} file inside a directory
listed in \tttt{:se runtimepath} will be loaded
(e.g. \tttt{.vim/plugin/something/ascript.vim}).
There are various ways to automate the installation
and enhance the management of plugins.
By default, one has the GETSCRIPT interface (see \tttt{:h getscript}),
that downloads latest scripts from Sourceforge as specified in \tttt{:h getscript-data},
and the Vimball interface, which creates and loads a Vimball for a
plugin.
Such a Vimball may be created with \tttt{:[range]MkVimball <filename> path}, where range specifies lines
that hold paths to files to be included in the \tttt{<filename>.vba} Vimball.
The Vimball can be installed in a system by \tttt{:source <filename>.vba}
or loading it at Vim startup with \tttt{\$ vim <filename>.vba}.

A Vim package is a directory that contain plugins.
It should be located inside a \tttt{pack/} directory
somewhere in the directories listed with \tttt{:se runtimepath}.
The plugins found in \tttt{pack/<packName>/*/start/} are loaded
at startup, the plugins found at
\tttt{pack/<packName>/opt/**} are loaded with\\
\tttt{:packdd <script\_or\_directory\_name>}.

All directories Vim looks for scripts are described
in \tttt{:h vimfiles} and are basically set by
\tttt{:se runtimepath} and conventions inside
each directory therein, such as to look for \tttt{vimrc} files
and \tttt{plugin/} or \tttt{pack/} directories.
Vim scripts can be loaded conditionally,
e.g. only if a function is used as in \tttt{:h autoload-functions}.
Example of such are filetype plugins (enabled by a \tttt{ftplugin/<filetype>.vim} file,
e.g. inside a plugin directory).
There is a number of plugin managers for Vim.
Pathogen and Vundle seem to be the most popular,
one because of its minimalism, the other because of
advanced features, e.g. for searching and installing plugins with
colon commands.

\subsection{Spell and spelllang (en and pt\_br)}
One might set the spelling language with\\ \tttt{:se spelllang=en\_us}
or \tttt{:se spelllang=pt\_br}
and toggle spell checking with \tttt{:setl spell!}.
Depending on the activities being performed,
these commands are used so often that one might use mappings as in~\cite{vimrc}.
Currently, Vim will download the files for a specific language if
not found in system.

\subsection{Scripting, Functions, Vimscript and other languages (e.g.  Python)}\label{script}
In Vimscript,
the colon commands (also Ex commands or command-line commands) are related through spaces,
punctuation and keywords (see \texttt{:h script}).
Scripting the Vim editor can also be accomplished using other languages,
as well documented e.g. in \tttt{:h python}.
Functions are defined through colon commands and are called
inside colon commands e.g. \tttt{:call MyFunction()}
or \tttt{:echo MyFunction(4)}.
Notice that functions are not commands but might be bind to them
through colon commands e.g. \tttt{nnoremap gF :call MyFunction()<CR>}.
Executing source files is very straightforward with
\tttt{:so <filename>.vim},
and one can always use the Ex mode for rapid scripting.
At the same time, the \tttt{:term} and terminal-job mode make
scripting other software more convenient, as output is
promptly navigated and copied (as discussed in Section~\ref{terminal}).

\subsection{Color}\label{visual2}
Newer versions of Vim support 24-bit true color (aka 16 million colors)
in terminal Vim (in gVim true color received support earlier).
The terminal must support true color, and tests are available e.g. in
\url{https://gist.github.com/XVilka/8346728}.
Then Vim needs to be set to use true colors with
\tttt{:set termguicolors}.
If using 8 or 16 bit colors, Vim uses the color palette from the
terminal, if using true color each color is defined directly.
Settings for using true color inside Byobu/Tmux involve tweaking
and are available in~\cite{vimrc}.
Good color schemes to use with true color are Gruvbox
and Solarize.
One might source syntax files at any time to change syntax, usually
though linking tokens to syntax groups as in
\tttt{:syntax keyword <group\_name> <token1> <token2> <token3> ...},
or
\tttt{:syntax match <group\_name>	<pattern>},
and then relating the group to another group: 
\tttt{:highlight link <group\_name> <group\_name2>}
or to group characteristics directly:
\tttt{highlight <group\_name> guifg=\#ffffff}.
If you change a syntax file, reloading a file with with \tttt{:e<CR>}
updates the highlighting on the window with the corresponding file type.
A complete syntax highlighting support typically involves at least three files:
\begin{itemize}
  \item \tttt{ftdetect/<filetype>.vim}, where the file type is
    detected with e.g.
    \tttt{:autocmd BufNewFile,BufRead *.<file\_extension> setfiletype <filetype>}
  \item \tttt{ftplugin/<filetype>.vim} with general settings for the
    file type, such as: 
    \tttt{:set tabstop=2 softtabstop=2 shiftwidth=2 textwidth=70
    expandtab autoindent}.
  \item A \tttt{syntax/<filetype>.vim} file, with bindings between
    tokens and highlighting groups; and highlighting group definitions.
\end{itemize}

This scheme is implemented very straightforward in this
plugin~\cite{tokipona}
for highlighting text in the Toki Pona language~\cite{tpLang}.
Syntax highlighting plugins are as file type plugins,
but also have a \tttt{syntax/<filetype>.vim} file
relating keywords, matches and regions to highlighting groups. 
One might see every highlighting group, and their final visual results,
with the commands
\tttt{:so \$VIMRUNTIME/syntax/hitest.vim}
or \tttt{:so hi}.
The \tttt{ftdetect/} and \tttt{ftplugin/} folders load as expected in the \tttt{plugin/}
directory, but \tttt{syntax/} files has to be moved to
\tttt{$\sim$/.vim/syntax/}, unfortunately.

\subsection{Fonts}
The fonts are defined by your terminal or inside gVim.
\ttt{C-+} and \ttt{C--} can be used to change font size.
Some settings for fonts, such as boldface, might be set using
the syntax highlighting facilities described in the previous section.

\section{Conclusions and further work}\label{conc}
This document seems reasonable as an overall reference of the Vim editor,
at least for my usage and level of proficiency.
Given the folkloric milestone of using Vim for 10 years,
this article might serve as a benchmark for one to relate
it's current use and understandings.
As a pedagogical material, it seems to be unique in the emphasis
on namespaces, understood as commands, variables, state-related lists, etc,
especially in Section~\ref{namespaces},
and the reference to the standard Vim documentation
to achieve the DRY KISS style described in Section~\ref{intro}.

\noindent {\bf Potential enhancements} to this document include:
\begin{itemize}
  \item The discussion of facilities such as reading emails and connecting over ssh.
    There is a working hack in~\cite{vimrc} for browsing over the WWW,
    but such aspect of Vim usage might receive more attention
    given that it is comfortable to navigate and edit in Vim
    and the resulting integrated environment.
  \item Updating of the information I can find about the issues discussed,
    such as about status lines in Section~\ref{visual1}.
  \item Include a discussion about Neovim.
    I have never used it, but it seems to be reaching a considerable user base
    and it might be feasible to give an account of Vim and Neovim
    after some tests and researching the official and unofficial documentation.
  \item Better cite documentation, plugins and Vim authors.
    I preferred to keep the references inline through \tttt{:h} commands
    and URLs, more in accordance with the style of Vim documentation,
    but bibliographical items constitute a valuable asset for academic literature,
    and some authors might find their work more respected if more
    thoroughly cited.
  \item An analysis of my usage, e.g. according to \url{http://www.drbunsen.org/vim-croquet/}.
    This undertake might benefit from data from many users,
    which favors the potential plugin for usage analytics described in the next bullet list.
\end{itemize}

\noindent {\bf Potential next steps} in using Vim include:
\begin{itemize}
  \item Measure the performance of text mining routines in Vimscript against those implemented in C or Python.
  \item Enhance the HTTP browsing capabilities of~\cite{vimrc}.
  \item Better integrate Python and Vimscript, especially for data visualization
    and syntax highlighting management,
    in accordance with the visualization issues described in
    Sections~\ref{visual1} and~\ref{visual2}.
\end{itemize}

\noindent This is a selection of the issues that might entail
plugins and that are more prominent for me:
    \begin{itemize}
      \item listing all the mappings available in each mode and the typing combinations which are available
        for new mappings.
        Maybe already group possibilities by criteria such as length
        of sequence and how central are the keys.
      \item Sessions, as described in Section~\ref{visual1}.
      \item AA messages (shouts): to keep track, document and share of working sessions
        as in~\cite{aa1,aa2}, with capabilities to manage AA sessions,
        send visual or sonic cues for temporal marks, use Vim state to build AA shouts,
        relate AA sessions to other media, such as software repositories,
        screencasts and images, interact with IRC channels and other social platforms.
      \item Slick Vim: a collection of the settings and mappings I use.
        Enhancements such as using \ttt{C-} commands also to browse tabs,
        shortcuts to join a window into a tab,
        and dummy minimal plugins as simplest possible, then file type
        and then syntax highlighting,
        Some more elaborate tweaks should also be present, such
        as breaking lines in sensible places for natural language texts
        while respecting e.g. \tttt{:se textwidth}.
      \item Dealing with .swo and .swp temporary files.
        In summary, if the restored .swo file has the same content
        and the file being opened,
        the restoration phase can be omitted.
        If the contents differ, Vim should open a tab with each file
        in a vertical split and run \tttt{:windo :diffthis}.
      \item Rendering images and equations.
        These are useful for using Vim in presentations
        or achieving a textual representation when it is mandatory,
        such as to comply with the limitations of a platform (e.g. Vim editor). 
        but also hold stylistic merits as ASCII art is often
        very appreciated.
        One can both obtain an ASCII representation of a binary image (e.g. JPG, PNG),
        and can directly render ASCII charts from data using cues such are shape, position
        and color.
      \item Redirecting the commands that usually show the results in
        a 'more' interface (e.g. status listings such as in
        Section~\ref{state}),
        which cannot be searched nor copied nor persists if one
        returns to editing a file.
        Ideally, it should parsed and linked quickfix or location
        window, and the syntax highlighting maintained.
        The basic idea is to use \tttt{:redir} command to redirect the output
        of such commands with \tttt{:se nomore}.
        Reasonable functions (and convenient commands) for having the
        output of such commands in a standard Vim window are
        in~\cite{vimrc}.
      \item Slide presentations. I've been using some automation for browsing
        slides and opening figures. Some Vim users asked for the settings and
        commands. They are very elementary use of registers which are executed over
        arbitrary but consistent textual patterns.
      \item For bringing back all the splits after an \tttt{:only}.
        Also for bringing back the tabs after a \tttt{:tabo}.
      \item Run Python excerpts from a file being edited in an IPython shell. 
        The buffer number of the terminal window should be stored,
        and than any selected lines should be run on that instance.
        Mappings should make available all movements to fetch the
        excerpts, execute registers, remain on script or in the IPython shell.
        Also, current Terminal-Job mode can be improved easily with
        mappings, such as \ttt{C-O} for one normal mode command.
      \item For keeping track of the usage and making analysis
        for optimizing the usage, as described in~\url{http://www.drbunsen.org/vim-croquet/}.
        Usage analytics.
      \item To facilitate the tweaking of syntax highlighting.
        This should include ways to easily access and change the
        syntax highlighting scheme and dump it to a file (dumping
        current highlighting scheme to a working Vimscript file
        is currently not supported by Vim!).
        Should also include changing the color scheme and highlighting scheme incrementally
        and selectively using the features described in Section~\ref{visual2}
        and in~\cite{tokipona}.
      \item Color schemes for true color.
        The standard colors (e.g. blue, elflord) loose some of the distinctions,
        e.g. SpellBad tokens are not highlighted on these color schemes
        if you have a functioning \tttt{se termguicolors} (standard GUI mode
        and available in terminals since recent versions of Vim 8,
        see Section~\ref{visual2}).
        In~\cite{vimrc} are some lines that make such color schemes over
        \tttt{:colorscheme blue}.
        This design of color schemes over the standard color schemes might be
        a very simple and effective way to make new color schemes.
        Such color schemes should also make use of the discussions
        in~\cite{tokipona}.
    \end{itemize}

\subsection*{Acknowledgments}
FAPESP (project 2017/05838-3); Vim developers and documentation maintainers;
Vim user community. 

\appendix
\section{Example of usage session}
I usually begin by opening a file or directory
with \tttt{\$ vim <filename>}.
The color scheme is alternated between
blue and GruvBox with
\tttt{:colo gruvbox} and \tttt{:colo blue}.
I open a vertical split and then move
the window to a new tab using
\tttt{:vs} and \ttt{C-W\_T}.
I then search for tokens related to
the enhancements I want to make or
the knowledge I want to acquire.
I go back to the previous tab with \tttt{gr}
and make a global replace with
\tttt{:\%s/<this>/<that>/g}.
On adding dots to sentences,
I record in the \tttt{"q} register
the sequence \tttt{jA.jj},
and use it as a macro 10 times by
\tttt{10@q}.
I move to the other tab with \tttt{gt}
and open a terminal window with \tttt{:term}
for compiling latex files.
I start another terminal with \tttt{<C-W> :term}
for opening the resulting PDFs with evince.
If any new idea comes to mind and I have time,
\tttt{\textbackslash\textbackslash s} opens 
my vimrc~\cite{vimrc} for editing and \tttt{\textbackslash s}
sources it.
If there is e.g. code or notes in other projects I am working
on, I reach them through \tttt{<Alt-arrows>} because they
are in separate Vim instances
in Byobu/Tmux sessions and windows.
Because I stay for hours editing (\LaTeX and Python
scripts), I change the background to red, and eventually to green.
Because the blue color scheme does not highlight the spelling errors
if using true color, I run
\tttt{:hi SpellBad guifg=red guibg=lightblue}
to see the words found wrong by the spell checker.

\section{My vimrc file and usage}
In using Vim with my vimrc file~\cite{vimrc},
I mostly toggle the status line with \texttt{\textbackslash\textbackslash B}
and the tab line with \texttt{\textbackslash\textbackslash T}.
Save and close windows with \texttt{\textbackslash w} and \texttt{\textbackslash q}.
The mappings for transitioning through splits and tabs
are also used constantly.
Although the file is commented, one should
look for the help pages on the options that (s)he does
not understand, as a thorough explanation
of the file is tedious and out of the scope
of this document.

\section{Example notes on mappings}\label{notes}
\texttt{:h index} shows all the default mappings
while \tttt{:map} shows the user-defined mappings.
By considering such information, one can make
useful observations exemplified in this Appendix.

\subsection{Normal mode}
Every letter and character in the keyboard is used.
In Normal mode: \ttt{TAB} is the same as \ttt{C-I},
\ttt{BS} and \ttt{C-H} are the same as h.
\ttt{C-J} and \ttt{C-N} are the same as j.
\ttt{C-P} is the same as k.
Space is the same as l.
\ttt{C-[} and \ttt{Esc} are not used,
\tttt{<C-\textbackslash> a-z} are reserved for extensions,
and \ttt{C-\_} is not used.
\tttt{+} is the same as \ttt{CR} and they are both not very useful.
Del is the same as x.

Many \tttt{][} combinations are not used, e.g.
with \tttt{abhjklfg}.
The \tttt{\_} command might be used as a more powerful \tttt{\^},
leaving it free for mappings.

Directions, home, end, page up and down, insert, all have mappings
in more centraly located keys.
The letters beggin unbounded mappings (require another character to
trigger an action without having to wait for \tttt{:se timeoutlen}): g,z,[].

The \ttt{C-(HJKL)} commands are redundant,
with the exception of \ttt{C-L} which redraws the screen,
so it is a reasonable choice to use them to move focus
of the editor to splits in the \tttt{hjkl} directions.

Many key combinations are available for new mappings through the \tttt{g}
and \tttt{z}, and \tttt{v} commands.
They have typical uses, e.g. \tttt{z} for folds,
spell checking and some movements (mainly when wrap is set).

\subsection{Insert mode}
All standard character keys are used for entering text.
\tttt{<C-G>(j,k)} can be achieved by \tttt{<C-O>(j,k)} which is more powerful
in moving through multiple lines.
\ttt{C-$[$} is the same as \ttt{ESC}.
\ttt{C-J} and \ttt{C-M} are \ttt{CR}.
\tttt{<C-\textbackslash> a-z} are reserved for extensions,
other combinations with \ttt{C-\textbackslash} are not used.

\section{My \tttt{:version}}
I should keep the output of \tttt{:version} executed on the system
in which I wrote this document in this link:
\url{https://github.com/ttm/vim/raw/master/version.txt}.
It was compiled with this Makefile~\cite{makefile}
in a 16.04 LTS Ubuntu Linux,
and is tagged as 8.0, Included Patches 1-1173.

\section{Key notation and meaning for Vim}\label{sec:not}
The tokens \ttt{C-X}, \ttt{S-X} and \ttt{M-X}
mean \tttt{Ctrl+x}, \tttt{Shift+x} and \tttt{Alt+X}
(or \tttt{Meta+x}).
\ttt{A-X} is the same as \ttt{M-X} and refer to
Alt or Meta keys.
\ttt{C-X}, \ttt{C-S-X}, \ttt{C-S-x} and \ttt{C-x} are the same,
i.e. Vim does not distinguish between lower and uppercase 
letters in \ttt{C-} commands.
The \ttt{M-x} and \ttt{M-X} (or \ttt{M-S-x} or \ttt{M-S-X})
are different commands,
and they are not used by Vim's builtin keys.
Such \ttt{M-} key combinations potentially conflict
with shortcuts of other programs (e.g. \ttt{M-F10} for a terminal menu),
but are otherwise a large and safe set of combinations
for one to use.

In summary,
one might choose mappings that overwrite default key combinations,
use combinations not used by Vim (e.g. some of the \tttt{g} and \tttt{z}
normal commands, \ttt{C-} and \ttt{S-} commands, or any \ttt{M-} and
\ttt{M-S-} command), or
create new mappings using \tttt{<leader>} and \tttt{<localleader>}.
More information about key notation is in \tttt{:h key-notation}.
Default mappings are in \tttt{:h index} and user-defined (and plugin-defined) mappings
in \tttt{:map}.

\end{document}